\documentclass[aps,prl,preprintnumbers,twocolumn,superscriptaddress,showpacs,floatfix]{revtex4}
\usepackage{epsfig}
\usepackage{dcolumn}
\usepackage{bm}
\usepackage{amssymb}

\begin{document}

\title{Elliptic Flow Suppression due to Hadron Mass Spectrum}
\author{Jacquelyn Noronha-Hostler}
\author{Jorge Noronha}
\affiliation{Instituto de F\'{\i}sica, Universidade de S\~{a}o Paulo, C.P. 66318,
05315-970 S\~{a}o Paulo, SP, Brazil}
\author{Gabriel S.\ Denicol}
\affiliation{Department of Physics, McGill University, 3600 University Street, Montreal,
Quebec, H3A 2T8, Canada}
\author{Rone P.~G.~Andrade}
\affiliation{Instituto de F\'{\i}sica, Universidade de S\~{a}o Paulo, C.P. 66318,
05315-970 S\~{a}o Paulo, SP, Brazil}
\author{Fr\'ed\'erique Grassi}
\affiliation{Instituto de F\'{\i}sica, Universidade de S\~{a}o Paulo, C.P. 66318,
05315-970 S\~{a}o Paulo, SP, Brazil}
\author{Carsten Greiner}
\affiliation{Institut f\"ur Theoretische Physik, Goethe Universit\"at, Frankfurt, Germany}

\begin{abstract}
Hadron resonance gas models provide a good description of the equation of state of quantum
chromodynamics determined by lattice QCD calculations at temperatures $T
\sim 100-155$ MeV. In this paper we investigate the effects of an
exponentially increasing hadron mass spectrum (Hagedorn spectrum) on the azimuthal anisotropy of the
rapidly expanding matter formed in ultrarelativistic heavy ion collisions.
If the temperature at which the conversion from fluid degrees of freedom to hadrons is sufficiently close to the Hagedorn temperature, the
production of Hagedorn resonances suppresses the differential elliptic flow of all hadron species.
\end{abstract}

\maketitle



Ultrarelativistic heavy ion collisions at the Relativistic Heavy Ion
Collider (RHIC) and the Large Hadron Collider (LHC) are able to reach
temperatures high enough to create and study the Quark-Gluon Plasma (QGP). Nevertheless, extracting the QGP
properties from heavy ion collisions is still a challenge since this state
of matter is only created transiently. Experimentally, it is only possible
to measure hadrons, leptons, and photons produced throughout the collision
with most of the hadrons being formed in the final stages of the collision.

Therefore, in order to study the properties of the QGP, we have to model the
whole heavy ion collision: from the formation and thermalization of the QGP
to the dynamics of the hadron rich medium formed at the end of the
reaction. The theoretical modeling of ultrarelativistic heavy ion collisions
has become very sophisticated in the last decade, achieving a level of
precision without precedent. Currently, a state of the
art description of the QGP formation and subsequent evolution should incorporate: 1) a description of the pre-equilibrium
phase of the QGP using gluon saturation inspired models \cite%
{iccgc}, 2) the time evolution of the QGP using second order relativistic
dissipative fluid dynamics, 3) a conversion to a hadron rich phase in local
thermodynamic equilibrium via Lattice QCD (LQCD) equations of state \cite%
{Aoki:2006we,Borsanyi:2010cj,fodormore,Bazavov:2011nk,Huovinen:2009yb}, and finally 4) a description
of hadronic matter using hadronic cascade simulations. For the current theoretical description of heavy ion collisions see \cite%
{song,Hirano:2005xf,Niemi:2012aj,Bozek:2012fw,Huovinen:2001cy,Nonaka:2006yn,Werner:2010aa,Soltz:2012rk,romatschke,gabrielshear,bjorn,nexspherio,hanna,Bass:1999tu,Teaney:2000cw}.

A simulation that includes at least items 2) and 4) is usually referred to
as hybrid model (see, for instance, \cite{song}), which couples fluid
dynamics to hadronic transport simulations. These are thought to provide a
more reliable description of the hadronic matter formed at the late stages
of the collision which would then remove uncertainties in the extraction
of thermodynamic and transport properties of the QGP. 

In practice, in hybrid models the transition between the fluid degrees of freedom to the hadronic ones is implemented via the Cooper-Frye method \cite{Cooper:1974mv}. Usually this procedure employs an isothermal space-time hypersurface, with the distribution of hadrons computed in such a hypersurface being used as an initial condition and also boundary condition for the hadron cascade simulation. Note that in order to implement this process it is necessary to provide not only the momentum distribution of hadrons inside each fluid element but also the hadron chemistry of the system. At the temperatures when most hybrid models switch from fluid dynamics to transport theory, $T_{sw}\sim 130-165$ MeV, the hadron chemistry is not fully known and is still a subject of intense investigation. 


Recently, LQCD thermodynamics \cite{Borsanyi:2010cj} at temperatures $T\sim 100-155$ MeV has been shown to be compatible with calculations performed using Hadron Resonance Gas (HRG) models where 
the hadron density of states increases exponentially $\rho (m)\sim g(m)\exp(m/T_{H})$ with $\lim_{m/T_H \gg 1}g(m)=0$ \cite{Majumder:2010ik,NoronhaHostler:2012ug}. The main parameter that characterizes 
the increase of the hadron density of states in this case is $T_H$, known as the Hagedorn temperature \cite{Hagedorn:1965st,Frautschi:1971ij}, which is an energy scale of the order of the QCD (pseudo) phase 
transition temperature \cite{Aoki:2006we,Borsanyi:2010cj,fodormore,Bazavov:2011nk}. In this paper, the hadronic states with masses larger than those measured in the particle data book \cite{Eidelman:2004wy} $m>2.5$ GeV, whose existence is implied by $\rho(m)$, are called Hagedorn resonances. These massive states were shown to affect chemical equilibration times \cite{NoronhaHostler:2007jf,NoronhaHostler:2009hp,NoronhaHostler:2009cf}, particle ratios \cite{NoronhaHostler:2009tz}, and the shear viscosity of a hadronic gas \cite{NoronhaHostler:2008ju,pal}. While an experimental confirmation of an exponential increase in the number of hadron states may be challenging \cite{Broniowski:2000bj,Broniowski:2004yh,Cohen:2011cr}, the validity of such an exponential spectrum is motivated by the success of hadron models in describing low $T$ lattice data in $SU(3)$ pure glue \cite{Meyer:2009tq,Borsanyi:2012ve} and also QCD \cite{Majumder:2010ik,NoronhaHostler:2012ug}. 

In this paper we show that the presence of Hagedorn resonances in the equation of state has an effect on the basic dynamical observables of heavy ion collisions 
leading to an increase in the total hadron particle spectrum and to a suppression of the differential elliptic flow of all hadrons species. The effect will be
more significant in hybrid models in which the fluid degrees of freedom are usually
converted into hadronic ones at isothermal hypersurfaces with temperatures where Hagedorn resonances are highly
populated. This introduces a new source of uncertainty that must be dealt with to correctly extract the value of the shear viscosity in the QGP formed in ultra-relativistic heavy ion collisions.

\begin{figure}[ht]
\centering
\includegraphics[width=0.4\textwidth]{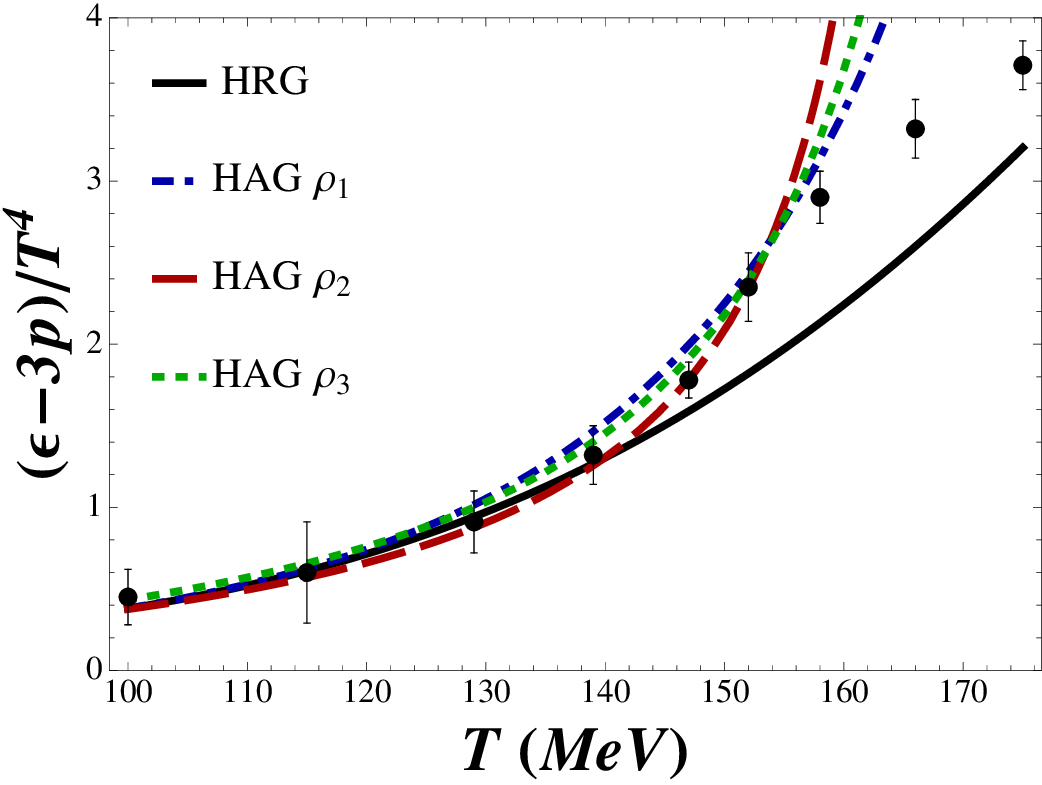}
\caption{(Color online) Trace anomaly in the hadron resonance model. The black solid curve denotes the result obtained in the standard HRG model. The dotted-dashed blue curve was computed using the model defined by the exponentially rising hadron mass spectrum $\rho_1(m)$, the long dashed red curve was computed using $\rho_2(m)$ while $\rho_3(m)$ was used to obtain the dotted green curve. The data points correspond to $N_t$ = 10 lattice data \cite{Borsanyi:2010cj}.}
\label{fig1}
\end{figure}

The thermodynamics of a resonance gas with Hagedorn-like resonances has been
studied a long time ago \cite{rafelski,kapustaolive,Welke:1990za,Venugopalan:1992hy}. The equilibrium pressure of
such a system at temperature $T$ is given by (we use classical statistics
throughout this paper, for simplicity) 
\begin{equation}
p(T)=\frac{T^{2}}{2\pi ^{2}}\int_{m_{min}}^{M_{max}}dm\,m^{2}\,\rho
(m)K_{2}\left( m/T\right) ,
\end{equation}%
where the integral is limited from below by a mass scale $m_{min}$ (taken
here to be zero) and from above by $M_{max}$. The standard
HRG pressure computed using the discrete set of hadron states from the
particle data group can be obtained from this continuum model by taking the
appropriate discrete limit of the integral above. The effect of Hagedorn resonances on the thermodynamics can be seen in Fig.\ \ref{fig1}. Following \cite{Majumder:2010ik,NoronhaHostler:2012ug}, the trace anomaly obtained from lattice calculations \cite{Borsanyi:2010cj} is compared to that of the HRG model and also to that found in Hagedorn gas models with density of states: $\rho_{1}(m)=A_{1}e^{m/T_{H1}}$ with $A_{1}=2.84$ GeV$^{-1}$ and $T_{H1}=0.252$ GeV, $\rho_{2}(m)=A_{2}e^{m/T_{H2}}/\left( m^{2}+m_{0}^{2}\right) ^{3/2}$ \cite{NoronhaHostler:2012ug} where $A_{2}=0.37$ GeV$^{2}$, $T_{H2}=0.167$ GeV, and $\rho_3(m)=A_{3}e^{m/T_{H3}}/\left( m^{2}+m_{0}^{2}\right) ^{5/4}$ where $A_3=0.63$ GeV$^{3/2}$ and $T_{H3}=0.180$ GeV \cite{NoronhaHostler:2012ug} and $m_{0}=0.5$ GeV (with maximum masses taken to infinity). One can see that the inclusion of Hagedorn states into the resonance gas model allows for agreement with lattice data up to $T \sim 155$ MeV \cite{Majumder:2010ik,NoronhaHostler:2012ug,footnote2}. 

In a single component gas, an increase in the system's total energy leads to an increase of the
individual energy of the particles and, consequently, to a higher
temperature and larger particle number density. However, in the case of a gas with Hagedorn resonances, any extra energy given to the system is used to produce more and more species of heavier particles (according to the
exponential spectrum), and not to increase the energy of each individual
particle species (or the overall system's temperature) \cite{Blanchard:2004du}. Since the typical switching temperature used in realistic hybrid models is similar to the Hagedorn temperature, this unusual
way to redistribute energy via the production of heavier resonances will
affect the overall momentum anisotropy of the particles that are
introduced in the hadronic transport codes.

In fact, consider the standard Cooper-Frye procedure \cite%
{Cooper:1974mv} for an ideal fluid over an isothermal hypersurface $\Sigma $
with temperature $T_{sw}$. For simplicity, we take the perfect fluid approximation in this first study to better illustrate the effects of Hagedorn resonances. 
%
%
The particle distribution per unit rapidity for a given species of mass $m_{a}$ and degeneracy $g_{a}$ is given by 
\begin{equation}
\frac{dN_{a}}{dyp_{T}dp_{T}d\phi }=\frac{g_{a}}{(2\pi )^{3}}\int_{\Sigma
}d\Sigma _{\mu }p^{\mu }\,e^{-p_{\mu }u^{\mu }/T_{sw}},
\end{equation}%
where $p_{T}$ is the particle's transverse momentum, $\phi $ is the momentum
azimuthal angle, $u^{\mu }$ is the fluid's 4-velocity, and $p^{\mu }$ is the
on-shell 4-momentum of the particle. The total particle distribution is
given by sum over all the states produced at the freeze-out surface $%
dN/dyp_{T}dp_{T}=\sum_{a}dN_{a}/dyp_{T}dp_{T}$. Approximating this sum by an
integral over the density of states $\rho (m)$, we then obtain that the
overall particle distribution per unit rapidity is 
\begin{equation}
\frac{dN}{dyp_{T}dp_{T}d\phi }=\int_{0}^{M_{max}}\frac{dm\,\rho (m)}{(2\pi
)^{3}}\int_{\Sigma }d\Sigma _{\mu }p^{\mu }\,e^{-p_{\mu }u^{\mu }/T_{sw}}.
\end{equation}%
The total particle spectrum is given by $\frac{dN}{dyp_{T}dp_{T}}=\int d\phi 
\frac{dN}{dyp_{T}dp_{T}d\phi }$. The elliptic flow coefficient per unit of rapidity of the hadrons emitted from
this hypersurface can be computed using the standard event plane method \cite{Poskanzer:1998yz} 
\begin{equation}
v_{2}(p_{T})=\frac{\int_{0}^{2\pi }d\phi \,\frac{dN}{p_{T}dp_{T}d\phi }\cos
[2(\phi -\psi _{2})]}{\frac{dN}{dyp_{T}dp_{T}}},  \label{v2}
\end{equation}%
where $\psi _{2}$ is the event plane angle. The integrated $v_2$ can be computed accordingly. The relevant $p_{T}$ range for
hydrodynamical behavior in ultra relativistic heavy ion collisions is $%
p_{T}^{max}< 3$ GeV. Very heavy resonances with transverse masses $%
m_{T}=\sqrt{p_{T}^{2}+m^{2}}$ a few times larger than $p_{T}^{max}$
contribute to make the overall distribution more isotropic since $e^{-p_{\mu }u^{\mu
}/T_{sw}}=e^{-m_{T}\gamma (1-\vec{p}_{T}\cdot \vec{v}/m_{T})/T_{sw}}\sim
e^{-m\gamma /T_{sw}}$. In fact, it has been known for quite some time that for heavy hadrons the differential elliptic flow generally increases slower with $p_T$ in comparison to that found for light hadrons \cite{Huovinen:2001cy,Borghini:2005kd}. Therefore, as $T_{sw}$ is brought closer and closer to $T_H$, more of these heavy states are emitted and this \textquotedblleft isotropization" mechanism induced by heavy resonances should lead to a suppression of the
overall differential elliptic flow of the matter. 

Moreover, note that as we increase $M_{max}$, more states are produced and the total number of particles increases. The $p_T$ spectrum is also enhanced and this effect becomes more significant at high $p_T$. This occurs because the exponential term $e^{m/T_H}$ in the density of states compensates the Boltzmann factor $e^{-m_{T}\gamma /T_{sw}}$ for very heavy states and one obtains considerably more particles in the spectrum at high $p_T$ by increasing $M_{max}$ (heavy particles should have flatter $p_T$ spectra in comparison to light particles). Also, while the velocity field in the hydrodynamic calculation is not particularly sensitive to the change in the EOS due to Hagedorn resonances, note that conservation of energy and momentum through the isothermal hypersurface implies that these heavy states must be produced when converting the fluid degrees of freedom into hadrons if the switching temperature is around 155 MeV. 


We tested these arguments by computing the elliptic flow coefficient of hadrons
emitted from an isothermal hypersurface of temperatures, $T_{sw}=130$ and $155$ MeV. The
isothermal hypersurfaces were computed by solving (boost invariant) relativistic ideal fluid dynamics.
We used a single averaged optical Glauber initial condition \cite%
{icglauber} to describe RHIC's 20-30\% most central events at $\sqrt{s}=200$ GeV \cite{footnote0}. 
We further assumed that the initial transverse flow of the system is zero. 
This will be sufficient to understand the effects of the
Hagedorn spectrum on the particle spectrum and elliptic flow, although
event-by-event simulations would be required to investigate higher order
Fourier coefficients. The equations of boost invariant ideal hydrodynamics
are solved for this initial condition using a Smoothed Particle Hydrodynamics (SPH) algorithm (a viscous version of this code has appeared in \cite{footnote}).  
In this first study, particle decays are not included and we fix the value of the
energy density at the initialization time (1 fm/c) to match the expected number of
direct pions for a given choice of the switching temperature. The equation
of state used in this calculation is the one presented in \cite%
{Borsanyi:2010cj}, whose low temperature behavior was shown to be compatible with a hadron resonance gas that includes a Hagedorn spectrum \cite{Majumder:2010ik,NoronhaHostler:2012ug,footnote1}.

In Figs.\ 2 and 3 we show the differential elliptic flow of all hadron species at isothermal hypersurfaces of
temperatures $T_{sw}=0.130$ GeV and $T_{sw}=0.155$ GeV, respectively. We assume that $T_{sw}=0.155$ GeV is the largest temperature at which one can still reliably say that the Hagedorn gas describes the lattice data. We used the previously defined density of states $\rho_1(m)$, $\rho_2(m)$, and $\rho_3(m)$. In both plots, the solid black curve corresponds to the value of elliptic flow in the case of an ordinary HRG without Hagedorn resonances while the short-dashed blue curve, the long-dashed red curve, and the dotted green curve correspond to the values of $v_2$ computed including Hagedorn resonances according to $\rho_1$, $\rho_2$, and $\rho_3$, respectively. The maximum mass of the Hagedorn resonances was taken to infinity although we have verified that the results already saturate when $M_{max} \sim 10$ GeV. 

One can see in both plots that the addition of Hagedorn resonances leads to a suppression of $v_2(p_T)$ with respect to the HRG calculation. Also, note that the $v_2$ computed using $\rho_2$ is smaller than that computed using $\rho_1$ or $\rho_3$, which was expected since $T_{H2}<T_{H3}<T_{H1}$. In fact, note that the elliptic flow suppression for $T_{sw}=0.155$ GeV is larger than that obtained when $T_{sw}=0.130$ GeV. This confirms our expectation that the production of heavy Hagedorn resonances leads to a suppression of elliptic flow if the switching temperature is sufficiently close to the Hagedorn temperature. We checked that the suppression does increase even further if $T_{sw}$ is taken to be $0.165$ GeV. Also, we verified that the spectrum becomes flatter due to the effect of resonances in the $p_T$ range where the suppression of $v_2(p_T)$ is more pronounced. 

The differential elliptic flow suppression discussed here can be of the same order of the typical differential elliptic flow reduction obtained due to the inclusion of $\eta /s\sim 1/4\pi $ viscous effects (see, for instance, \cite{bjorn}). It would be interesting to investigate the interplay between the suppression of elliptic flow induced by heavy resonance production and that coming from viscous hydrodynamic effects. 

Furthermore, we have checked that the inclusion of Hagedorn states in the calculation of the integrated $v_2$ for this centrality class leads to a change of only $4\%$ with respect to the value found for the HRG (we used $T_{sw}=0.155$ GeV) for the different parametrizations of the hadron spectrum considered here. This small change in the integrated $v_2$ is expected since the differential $v_2$ with Hagedorn state effects only starts to appreciably differ from that of the HRG when $p_T > 1.5$ GeV and that region in transverse momenta contributes very little to $p_T$ integrated quantities. 

We note that the hadrons and resonances emitted from the isothermal hypersurface rescatter and also decay, leading to  changes in the momentum distribution of hadrons and its anisotropy. This effect is not included in this work. However, one would expect that particle decays should not enhance the overall anisotropy of the expanding matter. Nevertheless, the most appropriate way to study the effect of heavy resonance dynamics and decay would be to include them in the current hadronic cascade simulations. In order to do so one would need to know the mass, the quantum numbers, and the cross sections of these Hagedorn states, which are not yet known. A possible way to determine these quantities was proposed in \cite{Pal:2005rb,Beitel:2014kza}. However, while hadronic transport models such as UrQMD do include multi-hadronic decays, the reverse process involving the formation of Hagedorn states in the HRG is not yet included. In fact, the biggest challenge to implement Hagedon states in transport codes may be the lack of detailed balance for this type of multi-particle processes. Thus, given these uncertainties, in this paper we chose to not include the effect of Hagedorn state decays into the elliptic flow. We hope to address these issues in the future.

\begin{figure}[ht]
\centering
\includegraphics[width=0.4\textwidth]{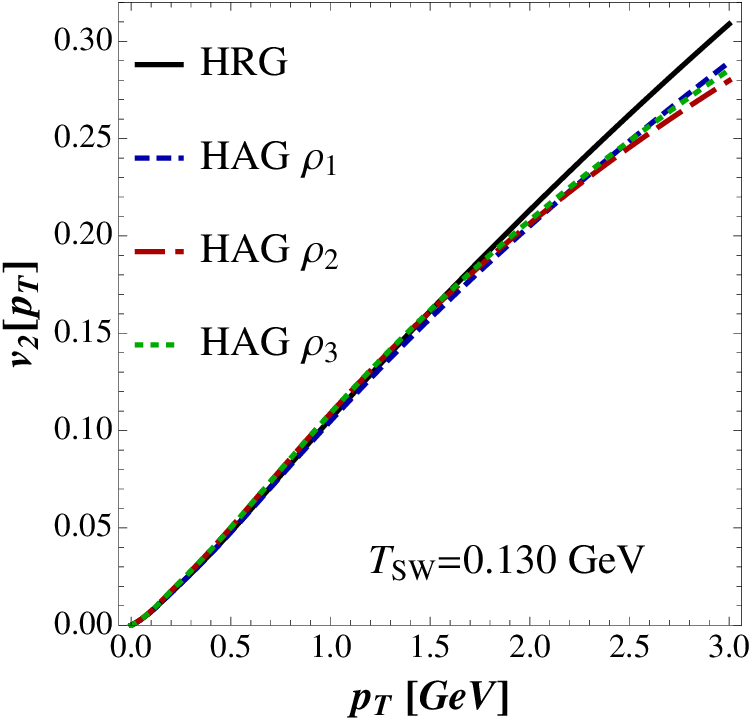}
\caption{(Color online) Differential elliptic flow coefficient of all hadron species for 20-30\% RHIC most central collisions computed using a single optical Glauber initial condition with freeze-out temperature $T_{sw}=0.130$ GeV (no particle decays are included). The solid black curve corresponds to the value of the elliptic flow in the case of an ordinary hadron resonance gas model including all the states in the particle data book while the short-dashed blue curve, the long-dashed red curve, and the dotted green curve correspond to the values of $v_2$ computed adding Hagedorn resonances that follow the density of states $\rho_1$, $\rho_2$, and $\rho_3$, respectively. The Hagedorn temperatures in $\rho_1$, $\rho_2$, and $\rho_3$ are $T_{H1}=0.252$ GeV, $T_{H2}=0.167$ GeV, and $T_{H3}=0.180$ GeV.}
\label{fig2}
\end{figure}

\begin{figure}[ht]
\centering
\includegraphics[width=0.4\textwidth]{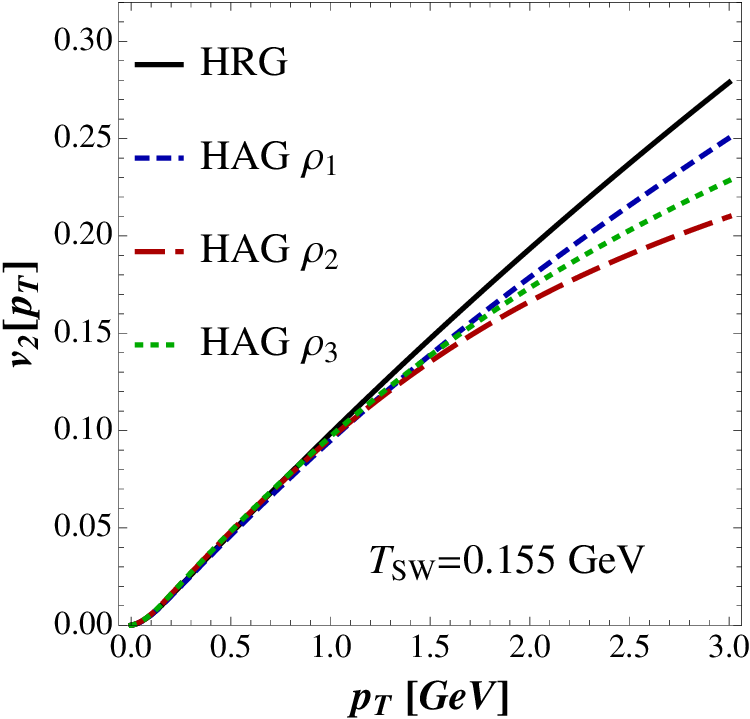}
\caption{(Color online) Differential elliptic flow coefficient of all hadron species for 20-30\% RHIC most central collisions computed using a single optical Glauber initial condition with freeze-out temperature $T_{sw}=0.155$ GeV (no particle decays are included). The solid black curve corresponds to the value of the elliptic flow in the case of an ordinary hadron resonance gas model including all the states in the particle data book while the short-dashed blue curve, the long-dashed red curve, and the dotted green curve correspond to the values of $v_2$ computed adding Hagedorn resonances that follow the density of states $\rho_1$, $\rho_2$, and $\rho_3$, respectively. The Hagedorn temperatures in $\rho_1$, $\rho_2$, and $\rho_3$ are $T_{H1}=0.252$ GeV, $T_{H2}=0.167$ GeV, and $T_{H3}=0.180$ GeV. A large suppression of $v_2$ with respect to the result from the HRG model is found when Hagedorn resonances are produced according to $\rho_2$ at $T_{sw}=0.155$ GeV, which is close to $T_{H2}$.}
\label{fig3}
\end{figure}

In conclusion, in this paper we showed that resonance production according to a Hagedorn spectrum 
leads to a significant suppression of the differential elliptic flow of all hadron species in ultrarelativistic heavy ion collisions if the switching temperature used in the conversion from fluid to hadronic degrees of freedom is close to the Hagedorn temperature. The isotropization mechanism implied by heavy resonance production should amount to a reduction of the higher flow harmonics as well, which can be verified by extending the study done here using event-by-event calculations \cite{nexspherio}. Our results indicate that the inclusion of Hagedorn resonances in the description of the hadron rich phase formed in heavy ion collisions may be relevant to improve the current estimates of the viscous effects in the QGP. 

This work was supported by Funda\c c\~ao de Amparo \`a Pesquisa do Estado de
S\~ao Paulo (FAPESP) and Conselho Nacional de Desenvolvimento Cient\'ifico e
Tecnol\'ogico (CNPq). G.~S.~Denicol is supported by the
Natural Sciences and Engineering Research Council of
Canada and thanks the University of S\~ao Paulo for
the hospitality provided during the completion of this work. C.~Greiner
thanks the Helmholtz International Center for FAIR within the framework of
the LOEWE program launched by the State of Hesse.



\begin{thebibliography}{999}
\bibitem{footnote0} Our average 20-30\% most central event corresponds to the average of 700 Monte Carlo Glauber generated events with number of participants between 139-196. At 130 MeV we assumed that 169 pions come from direct decays in the 0-5\% centrality class while at 155 MeV we obtain 110 direct pions.
\end{thebibliography}
\end{document}